\documentstyle[preprint,aps,epsf]{revtex}  

\begin{document}

\title{ Surface Structure and Catalytic $CO$ Oxidation Oscillations}
\author{R. Danielak $ ^{\mbox{\dag}}$, A. Perera, M. Moreau  \\
Laboratoire de Physique Th\'eorique des Liquides (*), \\
Universit\'e Pierre et Marie Curie,   \\
4, Place Jussieu, Tour 16, 5\`eme \'etage,  \\
75256 Paris, France, \\
\vspace{0.1in}
 M. Frankowicz \\ 
\dag Departement of Theoretical Chemistry, \\
Jagiellonian University, Cracov, Poland, \\ 
\vspace{0.1in}
 R. Kapral \\
Chemical Physics Theory Group, Department of Chemistry,  \\
University of Toronto, Toronto, Canada M5S 1A1. }

\maketitle

\footnotetext{Unit\'e associ\'ee au C.N.R.S }

\begin{abstract}
A cellular automaton model is used to describe the dynamics of the   
catalytic oxidation of $CO$ on a $Pt(100)$ surface. The cellular automaton
rules account for the structural phase transformations of the $Pt$
substrate, the reaction kinetics of the adsorbed phase and diffusion of
adsorbed species. The model is used to explore the spatial structure that 
underlies the global oscillations observed in some parameter regimes. The
spatiotemporal dynamics varies significantly within the oscillatory regime
and depends on the harmonic or relaxational character of the global
oscillations. Diffusion of adsorbed $CO$ plays an important role in the 
synchronization of the patterns on the substrate and this effect is also
studied.
\end{abstract}

\section{Introduction} \label{intro}
The study of catalytic oxidation processes on $Pt$ surfaces is
of considerable interest since the apparently simple kinetics 
of this reaction can
give rise to different types of behavior as a result of competing 
surface processes, substrate phase transformations and global 
coupling effects. Extensive investigations have been carried out 
in the last decade including both experiment and 
theory\cite{ertl1,ertlks}. Experiments have demonstrated the existence of a 
variety of non-linear phenonena such as
nucleation-induced front propagation, spiral wave dynamics,   
faceting structures and kinetic oscillations.\cite{ertlks} 

In this article we re-examine some aspects of the global oscillations 
observed in the oxidation of $CO$ on $Pt(100)$
surfaces.\cite{norton,ertl14} By ``global" 
we mean oscillations in concentrations of adsorbed species or $Pt$ 
substrate states averaged over the entire crystal surface or, in 
experiments, some macroscopic region of the surface. The principal
features that give rise to such oscillations, which involve coupling 
between surface phase transformations and the oxidation kinetics, 
were elucidated in earlier studies \cite{ertl14,imbihl,moller1,moller2} 
and are as follows:
The surface-catalysed $CO$ oxidation occurs by a Langmuir-Hinshelwood 
mechanism \cite{engel}:
\begin{eqnarray}
 CO+ \star  &{\mathrel{\mathop{\kern0pt 
 \rightleftharpoons}\limits^{k_1^{\alpha}}_{k_{-1}^{\alpha}}}}
 & CO_{ads}\;, \label{step1} \\
 O_2 + 2 \star &{\mathrel{\mathop{\kern0pt 
 \rightleftharpoons}\limits^{k_2^{\alpha}}_{k_{-2}^{\alpha}}}}
 &  2 O_{ads}\;, \label{step2} \\
\ CO_{ads} + O_{ads} & \stackrel{k_3^{\alpha}}{\rightarrow} 
& CO_2 \uparrow + 2 \star \;, \label{step3} \\
 CO + O_{ads} &  \stackrel{k_4^{\alpha}}{\rightarrow} &  
 CO_2 \uparrow +  \star \;, \label{step4}
\end{eqnarray}
where the $\star$ indicates an empty adsorption site. The first three steps 
(\ref{step1}-\ref{step3}) constitute the Langmuir-Hinshelwood mechanism.  
The fourth step (\ref{step4}) is included to prevent poisoning of
the $Pt$ surface by adsorbed oxygen, which is not observed in
experiment. It involves the direct reaction of an adsorbed oxygen atom
with $CO$ from the gas phase to produce $CO_2$ in the gas phase plus an 
empty surface site.

The actual mechanism of $CO$ oxidation on $Pt(100)$ involves a coupling
between the reconstruction of the $Pt$ surface and the adsorption and 
reaction kinetics\cite{ertl1}. The structure of the topmost layer of clean 
$Pt(100)$ differs from that in the bulk ($1 \times 1$) and is reconstructed 
to form a $hex$ structure.\cite{van} When the $CO$ coverage exceeds a critical 
value this reconstruction is lifted and the surface reverts to the $1 \times 1$ 
form. In general the reaction rates may be different for the two surface
phases and we have accounted for this possibility by appending the 
superscript $\alpha=1 \times 1$ or $hex$ to the rate constants in the 
Langmuir-Hinshelwood mechanism. Phase transformations between these
two surface forms are triggered by and, in turn, influence the characteristics 
of diffusion and reaction kinetics. It is the switch between two surface forms 
with high and low reactivities that is responsible for the existence of global 
kinetic oscillations.

Apart from these general observations on the origin of the kinetic oscillations, 
it is known that for $Pt(100)$ \cite{pt110}the spatial coupling of different parts 
of the surface is caused by the propagation of chemical fronts of oxidation across 
the crystal surface. \cite{ertl14,imbihl,moller1,moller2} The nature of the wave 
propagation processes depends on both the type of global oscillation and
the diffusion coefficient of adsorbed $CO$, $D_{CO}$. We investigate the 
spatial structure that underlies the global oscillations. 
 
The most extensive theoretical descriptions of wave propagation processes
associated with $CO$ oxidation on various $Pt$ surfaces have been 
based on reaction-diffusion equation descriptions of the spatiotemporal
dynamics \cite{diffeq}, but lattice-gas cellular automaton models 
\cite{moller1,moller2,ray1} 
have also been employed in such studies. These lattice-gas automata are
probabilistic cellular automata that employ a particle description of the
dynamics and should be distinguished from other cellular automata that are
designed to simulate reaction-diffusion equation
dynamics.\cite{rdca1,rdca2,rdca3} 
While most of the phenomena 
of interest occur on sufficiently long space and time scales that 
a continuum description using reaction-diffusion equations is appropriate, 
the cellular automaton models have the advantage that they naturally 
incorporate fluctuations that are responsible for the nucleation events that 
play a role in the process and, as well, can be extended to smaller scales where 
continuum models are no longer appropriate.
Cellular automaton models allow one to describe the system
at a mesoscopic level which incorporates the essential mechanistic features
of the reaction dynamics. Several such models for surface catalysed $CO$ oxidation  
employing simple Langmuir-Hinshelwood kinetics have been developed 
in the past years\cite{ziff,others}. 

In this paper we use a cellular automaton model for 
catalytic oxidation of $CO$ on $Pt(100)$ which accounts for 
the coupling among the local reaction kinetics, substrate phase 
transformations and diffusion of adsorbed species. This 
model is described in Sec.~\ref{CAmodel}. In Sec.~\ref{Oscres} the 
model is used to explore various aspects of the spatiotemporal dynamics in
the regime of global oscillations and, in particular, to investigate the
spatial structure that gives rise to these oscillations. Section~\ref{conc} 
contains the conclusions of this study.

\section{Cellular Automaton Model for $CO$ Oxidation} \label{CAmodel}
From the description given above we see that any model of this surface 
catalysed reaction must account for the coupling among phase transformations, 
diffusion and reaction. The cellular automaton model we employ has features 
that are like those of the Monte Carlo models\cite{ziff} in that asynchronous site updating 
rules are used, but there are also features in the model that are adopted from 
reactive lattice-gas automata \cite{pr} and, for the surface phase transformation 
process, rules similar to those of traditional cellular automata are 
used.\cite{ca} The dynamics is described by the following rules:
 
The cellular automaton dynamics is assumed to take place on a set of regular 
two-dimensional lattices ${\cal L}_{\tau}$. One lattice, ${\cal L}_s$, is used 
to represent the $Pt(100)$ substrate in its various phases 
($1 \times 1$ or $hex$). The dynamics of the adsorbed $CO$ and $O$ species 
take place on the ${\cal L}_a$ lattice.
Each node of ${\cal L}_a$ can support only one particle, either $O$ or
$CO$. The nodes of these two lattices have identical labels ${\bf r}$.

Adsorbed species at node ${\bf r}$ on ${\cal L}_a$ can react with molecules 
or atoms on neighboring sites ${\bf r}' \in {\cal N}_a({\bf r})$ where 
${\cal N}_a({\bf r})$ is some neighborhood of the node ${\bf r}$.
Similarly, we let ${\cal N}_s({\bf r})$ be the neighborhood of 
node ${\bf r}$ on the substrate lattice ${\cal L}_s$ which enters in
the description of the surface reconstruction process. Also, we define 
$\bar{{\cal N}}_{\tau}({\bf r})$ to be the node ${\bf r}$ plus its 
neighbors. In the
present work we consider square lattices with a total of $N$ nodes and 
take ${\cal N}_{\tau}({\bf r})$ to be  the Moore neighborhood consisting 
of the eight nearest and next-nearest neighbors of the node ${\bf r}$. 

The macroscopic rate constants defined in (\ref{step1}-\ref{step4}) are 
used to construct the microscopic reaction probabilities. We define 
an additional rate constant $k_{tr}$ that gauges whether a phase 
transformation will be attempted. For each such 
reaction rate constant $k_i$ we define the corresponding probability 
$p_i=k_i/k$ where $k=\sum_i k_i$. Thus, in the automaton rule described 
below $p_i$ will specify the probability with which any reaction 
in steps (\ref{step1}-\ref{step4}) or phase transformation are 
carried out at a lattice node.
 
There are three elements in the cellular automaton rule corresponding to 
the physical processes described above: reaction (to simplify terminology
we call adsorption and desorption ``reactions"),
phase transformation of the $Pt$ substrate and diffusion of adsorbed 
species. Each automaton time step consists of $N$ applications 
of the composition of the following operations:

A node ${\bf r}$ is picked at random. For node ${\bf r}$ a reaction 
$i$, including phase transformation, is chosen randomly with 
probability $p_i$. 

\vspace{0.1in} \noindent
{\bf Reaction}\\
If one of the reaction steps (\ref{step1}-\ref{step4}) is 
selected and the corresponding $i$-th reaction is 
sterically allowed, the reaction takes place. 
In the case of reactions involving simultaneously two nodes (for example, 
$O$ adsorption) the second neighboring lattice node is chosen 
randomly from ${\cal N}_a({\bf r})$ and the reaction is carried out 
if it is sterically allowed. The reaction probabilities account for 
the overcounting of identical pairs of nodes. 

\vspace{0.1in} \noindent 
{\bf Phase transformation process}\\
If the phase transformation step is selected then substrate 
phase transformations occur by the following
mechanism\cite{ray1}.  The substrate  can exist in two phases, 
$1 \times 1$ and $hex$. Consider the node ${\bf r}$
on ${\cal L}_s$ and let $n_{CO}^{\bar{\cal N}}({\bf r})= 
\sum_{{\bf r}'\in \bar{\cal N}_a({\bf r})} \theta_{CO}({\bf r}')$ be
the number of $CO$ molecules at node ${\bf r}$ plus its neighborhood. 
Here $\theta_{CO}({\bf r})=1$ if node ${\bf r}$ is occupied by $CO$ and 
zero otherwise. If we let $n$ be the number of neighbors in ${\cal N}_a$ 
then $\rho_{CO}^{\bar{\cal N}}({\bf r})=n_{CO}^{\bar{\cal N}}({\bf r})/(n+1)$ is the
fractional coverage of the region consisting of the node ${\bf r}$ and its 
neighborhood ${\cal N}_a$.
 
\begin{itemize}
\item
$1 \times 1$ domain formation mechanism: \\
if all nodes in $\bar{{\cal N}}_s({\bf r})$ are in the $hex$ phase and  
$\bar{{\cal N}}_a({\bf r})$ in ${\cal L}_a$ is covered by $CO$, then
the node ${\bf r}$ is changed to $1 \times 1$ with a probability $k_{H1}$. 

\item
$1 \times 1$ domain growth mechanism: \\
if node ${\bf r}$ on ${\cal L}_s$ is in the $hex$ phase 
and $\rho^{\bar{\cal N}}_{CO}({\bf r}) >  \rho_{max}$ then node ${\bf r}$ is 
changed
to $1 \times 1$ with a probability $F(n_{1 \times 1}) k_{G1}$, 
where $n_{1 \times 1}$ is the number of nodes in 
${\cal N}_s({\bf r})$ in the $1 \times 1$ phase. Here $\rho_{max}$ is some
lower fixed fractional coverage necessary for $1 \times 1$ domain growth.

\item
$hex$ domain formation mechanism: \\
if if all nodes in  $\bar{{\cal N}}_s({\bf r})$ are in the $1 \times 1$ phase 
and $\rho^{\bar{\cal N}}_{CO}({\bf r})=0$
then node ${\bf r}$ on ${\cal L}_s$ is changed to $hex$ with a probability 
$k_{1H}$.

\item
$hex$ domain growth mechanism: \\
if node ${\bf r}$ on ${\cal L}_s$ is in $1 \times 1$ phase and if 
$\rho^{\bar{\cal N}}_{CO}({\bf r}) <  \rho_{min} $,
then node is ${\bf r}$ changed to $hex$ with a probability 
$ F(n_{hex}) k_{GH}$, where $n_{hex}$ is the number of nodes in 
${\cal N}_s({\bf r})$ in the $hex$ phase. Here $\rho_{min}$ is some fixed
upper fractional coverage necessary for $hex$ domain growth.
\end{itemize}

We have introduced a function $F(n_{\alpha})$ ($n_{\alpha} > 0$)
of the neighborhood occupancy,  
$n_{\alpha}=\sum_{{\bf r}'} \theta_{\alpha}({\bf r}')$, 
$({\bf r}' \in {\cal N}_s({\bf r})$, $\alpha=hex,\; 1 \times 1$), where 
$\theta_{\alpha}({\bf r}')=1$ if ${\bf r}'$ is in phase $\alpha$ and zero 
otherwise. This function models the local structure and is
responsible for the phase transformation process. It can be used
in a phenomenological fashion to change the characteristics of this process. 
The form of $F$ depends on the microscopic details of 
the phase transformation process.\cite{ptnote}

\vspace{0.1in} \noindent
{\bf Diffusion}\\
No diffusion in the substrate is permitted. On ${\cal L}_a$ 
diffusion occurs through random jumps of the adsorbed species 
at node ${\bf r}$ to one of the empty nodes in ${\cal N}_a$. 
Some flexibility is allowed in the number of diffusion steps compared 
to reaction steps. One may pick additional nodes at random 
and carry out diffusive jumps. In this way large diffusion coefficients 
can be  simulated by increasing the number of diffusion steps per 
reaction step. 
Similarly very small diffusion can be simulated by increasing the number
of reaction steps for single diffusion step. A convenient 
variable to gauge the ratio of diffusion to reaction is  
$\delta=$(number of diffusion steps)/(number of
reaction steps).  In addition, the diffusion 
rule can be modified to account for the fact that the diffusion rates 
may be different 
on the $hex$ or $1 \times 1$ phases. Thus, the probability of a 
jump from node ${\bf r}$ may depend on the substrate state at that 
node. 

\section{Spatial Structure and Global Oscillations} \label{Oscres}
In order to study how reaction, diffusion and phase transformation 
can lead to global oscillatory behavior in $CO$ oxidation on $Pt(100)$ and 
to examine the underlying pattern formation that may accompany it, 
we have carried out a series of simulations of the automaton. 
In the results presented below we have taken 
$k_1^{1 \times 1}=k_1^{hex}$, $k_{-1}^{1 \times 1}=k_{-1}^{hex}
=0.001$, $k_2^{hex}=0$, 
$k_{-2}^{1 \times 1}=k_{-2}^{hex}=0$, $k_3^{1 \times 1}=k_3^{hex}=0.25$, 
$k_4^{1 \times 1}=k_4^{hex}=0.001$ and $k_{tr}=0.6$. These values 
reflect the fact 
that the major rate constant differences between the $hex$ and 
$1 \times 1$ phases occur in the $O_2$ adsorption on the surface: the 
sticking probability on the $hex$ phase is much smaller than that 
on the $1 \times 1$ phase. We have used $k_1^{1 \times 1}$ and 
$k_2^{1 \times 1}$ as bifurcation parameters.  
In the implementation of the phase 
transformation step we suppose $F(n_{\alpha})=n_{\alpha}^2$. 
The parameters that enter in the phase transformation steps are 
 $\rho_{max}=0.89$, $\rho_{min}=0.11$, $k_{1H}=k_{H1}=0.001$, 
$k_{GH}=k_{G1}=0.05$. $D_{CO}$  was varied by changing the value of 
the parameter 
$\delta$ defined in Sec.~\ref{CAmodel}. Typically $CO$ diffusion is 
much larger than that of $O$ and for simplicity in this study we have 
taken $D_O^{hex}=D_O^{1 \times 1}=0$ and 
$D_{CO}^{hex}=D_{CO}^{1 \times 1}=D_{CO}$ and investigated the spatial and 
temporal structure as a function of $D_{CO}$ and the kinetic parameters. 

Earlier investigations\cite{moller1,moller2} have shown that global 
oscillations on
$Pt(100)$ are found in a cusp-shaped domain in the ($p_{CO},p_{O_2}$) 
parameter plane. In our model these partial pressures are incorporated 
in the $k_1^{\alpha}$ and $k_2^{\alpha}$ rate coefficients. We have 
confirmed that our model reproduces this structure in the 
($k_1^{1 \times 1},k_2^{1 \times 1}$) parameter plane and, furthermore,
have examined how this bifurcation diagram \cite{binote} 
changes as $D_{CO}$ varies.
These results are presented in Fig.~\ref{pdiag} which shows the region of
oscillation for two values of the diffusion coefficient as gauged by
$\delta$; (a) is a large value of $D_{CO}$ with $\delta=10$ while 
(b) corresponds to a small value with $\delta=1/2$. The major qualitative 
feature one observes is that the oscillatory region persists to higher
$k_1^{1 \times 1}$ ($p_{CO}$) values when $D_{CO}$ is small. An
understanding of these ``phase" diagrams requires an examination of the
spatial structure that gives rise to the global dynamics. We now
characterize the global oscillations and then discuss the underlying
spatial structure. 

\subsection{Global oscillations}
\begin{figure}[htbp]
\begin{center}
\leavevmode
\epsfysize = 6in
\epsffile{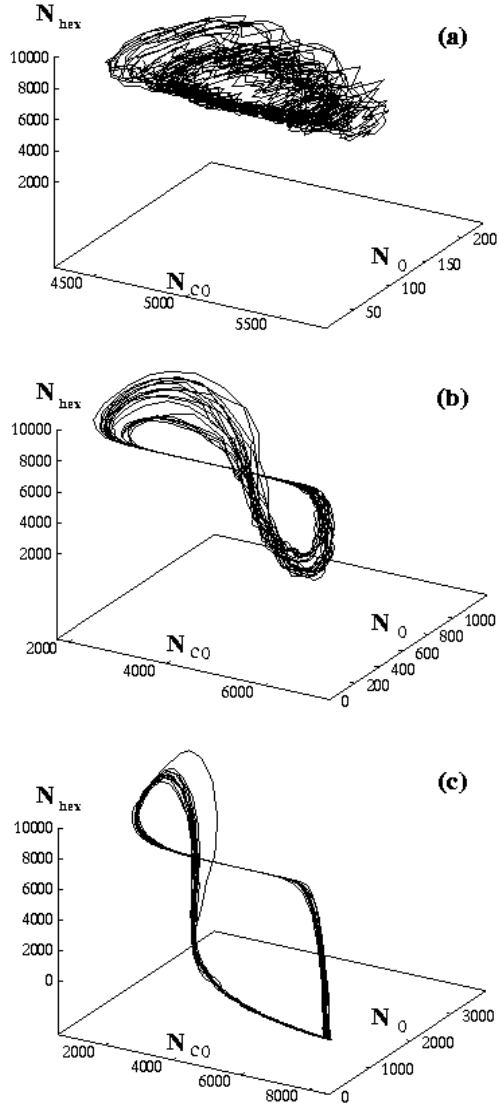}
\end{center}
\caption{Plots of the system trajectory for $\delta=10$ 
in the $(N_{CO},N_O,N_{hex})$ 
phase plane for $k_2^{1 \times 1}=0.04$ and (a) $k_{1}^{1 \times
1}=0.005$, (b) $k_{1}^{1 \times 1}=0.006$ and (c) $k_{1}^{1 \times
1}=0.0167$}
\label{gosc10}
\end{figure}

It is difficult
to precisely classify the nature of the bifurcations that are responsible
for oscillatory behavior since our cellular automaton model is a 
high-dimensional, particle-like model and we have no recourse to a simple
description based on ordinary differential equations (ode). Nevertheless, 
some general features of the bifurcation structure are consistent with those 
for the global oscillations on $Pt(110)$ described in Ref.\cite{ertl3}
where a two-variable ode model that did not include the details of the
spatial structure was studied. Figure~\ref{gosc10} presents phase space 
plots of the
total number of $CO_{ads}$, $O_{ads}$ molecules on the surface, $N_{CO}$ 
and $N_{O}$, respectively, and the total number of sites in the $hex$
phase, $N_{hex}$. (Clearly, $N_{1 \times 1}=N-N_{hex}$.) These three
variables define the phase space ($N_{CO},N_O,N_{hex}$) for spatially 
averaged dynamical
variables. This figure shows the phase space trajectory for these globally
averaged fields for fixed $k_2^{1 \times 1}=0.04$ and varying 
$k_{1}^{1 \times 1}$, moving from low to high values of this parameter.
The results suggest that the oscillations arise by a supercritical Hopf
bifurcation on the leftmost boundary of the oscillatory region since the
(noisy) global limit cycle is harmonic in character and 
its amplitude grows as
one moves away from the boundary towards higher $k_{1}^{1 \times 1}$ (cf.
panels (a) and (b) in this figure). This noisy periodic attractor
continues to grow in phase space size and evolves to the relaxation
oscillation depicted in	(c) close to the rightmost bifurcation boundary.
In this parameter region there are distinctly identifiable slow and 
fast portions to the cycle.
Note also that periodic attractor is far less noisy than in the harmonic
regime. When the bifurcation boundary is crossed this large noisy global
limit cycle suddenly crashes to a fixed point solution for the global
concentration variables. Such a sudden collapse to a fixed point solution
was found in ode models of oxidation on $Pt(110)$\cite{ertl3}. 

\begin{figure}[htbp]
\begin{center}
\leavevmode
\epsffile{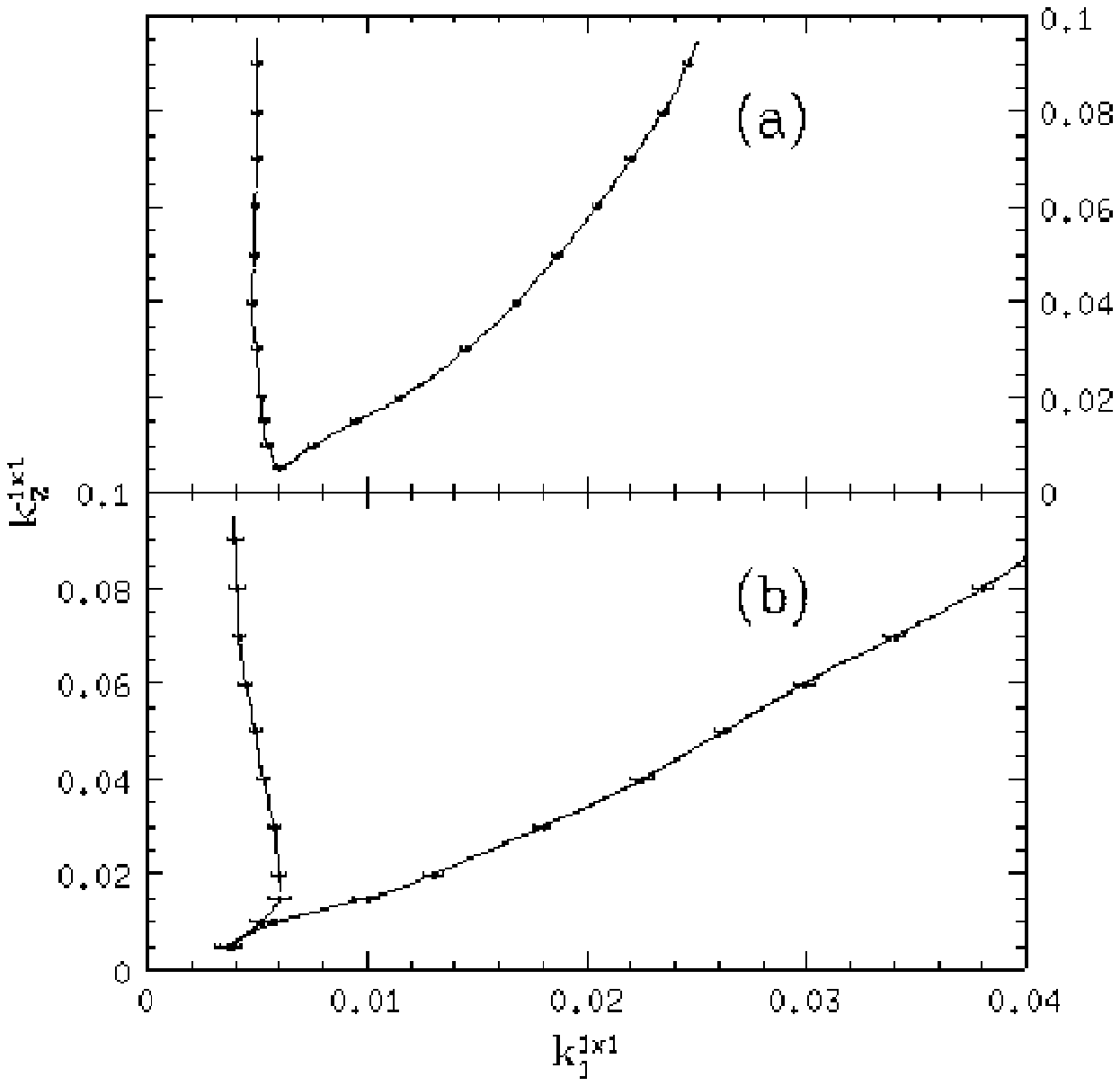}
\end{center}
\caption{Bifurcation diagram in the ($k_1^{1 \times 1},k_2^{1 \times 1}$) 
parameter plane showing the region of global oscillations for (a) 
$\delta=10$ and (b) $\delta=1/2$.}
\label{pdiag}
\end{figure}

\begin{figure}[htbp]
\begin{center}
\leavevmode
\epsfysize = 6in
\epsffile{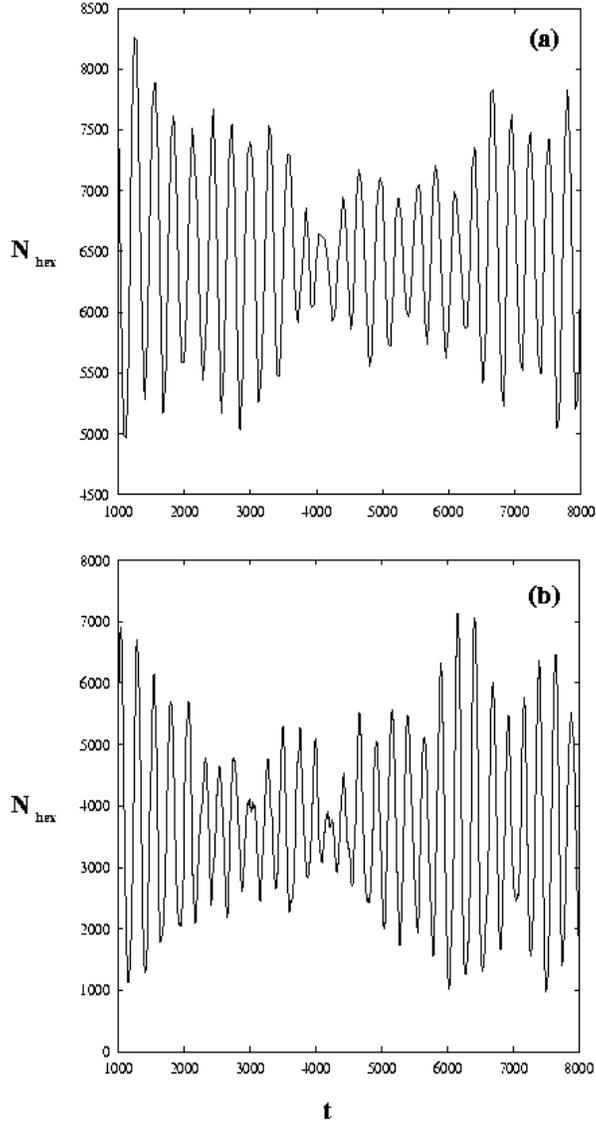}
\end{center}
\caption{Plots of $N_{hex}$ versus $t$ for $\delta=1/2$,  
$k_2^{1 \times 1}=0.04$ and (a) $k_{1}^{1 \times
1}=0.008$ and (b) $k_{CO}^{1 \times 1}=0.017$.}
\label{gosc.5}
\end{figure}

Figure~\ref{gosc.5} shows $N_{hex}$ versus $t$ for a small value of 
$D_{CO}$ ($\delta=1/2$) for two values of $k_{1}^{1 \times 1}$ along 
a line $k_2^{1 \times 1}=0.04$ in Fig.~\ref{pdiag}(b). These two values 
lie near the two bifurcation boundaries which showed harmonic and 
relaxation oscillations for large $D_{CO}$. The global
oscillations are highly irregular and much less well resolved, regardless 
of the location within the oscillatory regime. We have chosen to plot only
the time variation of a single variable rather than a phase plane plot
since the irregular character of the oscillations leads to a complex phase 
space trajectory where the oscillatory structure is obscured. As a result 
the bifurcation 
boundaries shown in Fig.~\ref{pdiag}(b) are subject to large uncertainties.  

\subsection{Spatial structure}
We now consider the spatial structure that underlies these global
oscillations. While it has been established that chemical waves
propagating across the surface are responsible for the existence of the
oscillations, we shall show that the character of these surface structures
varies considerably depending on the region of parameter space within the
oscillatory domain of the bifurcation diagram and the magnitude of
$D_{CO}$. 

\begin{figure}[htbp]
\begin{center}
\leavevmode
\epsfysize = 6in
\epsffile{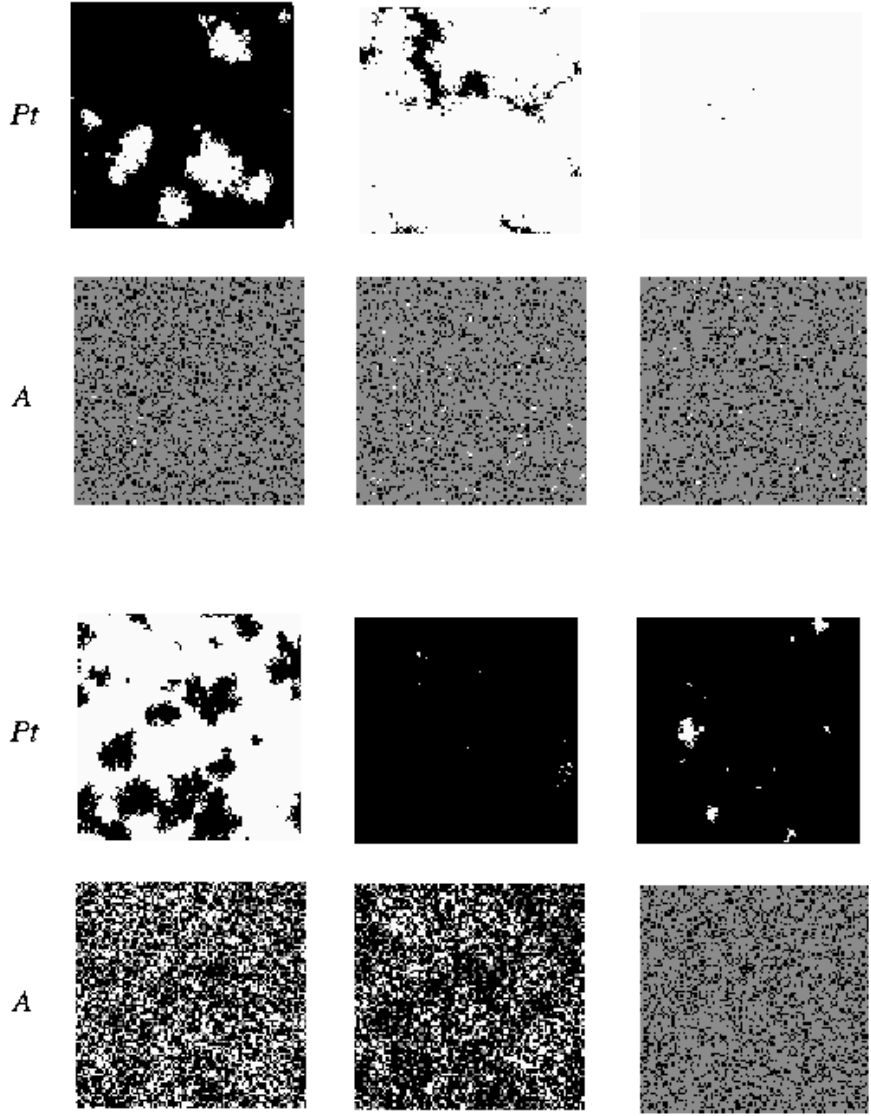}
\end{center}
\caption{Evolution of the $Pt(100)$ substrate and adsorbed phase in the regime of 
regular  relaxation oscillations for $k_2^{1 \times 1}=0.04$, 
$k_{1}^{1 \times 1}=0.015$ and $\delta=10$. On the $Pt$ surface labeled 
${\em Pt}$ the black areas correspond to the $hex$ phase and
the white areas to the $1 \times 1$ phase. The color coding 
for the species lattice labeled ${\em A}$ is as follows: black for $O$, 
gray for $CO$ and white for empty sites. Time increases from $t_1$ on 
the upper left to $t_6$ on the lower right. Starting with time $t_1$ the
subsequent times are: $t_2=t_1+60$, $t_3=t_1+90$, $t_4=t_1+420$,
$t_5=t_1+450$ and $t_6=t_1+630$.}
\label{spatd10r}
\end{figure}

The most regular global oscillations are found for large $D_{CO}$ in the
oscillatory region of the bifurcation diagram corresponding to high 
$k_{1}^{1 \times 1}$ where one observes relaxation oscillations. The
spatial pattern formation that accompanies these oscillations is shown in
Fig.~\ref{spatd10r} where the states of both the $Pt$ surface and the
adsorbed phase are shown for a series of six times within one global
oscillation period.  The upper panels in each time group (labeled 
${\em Pt}$) show the $Pt$ surface phases while the lower set of panels in 
each time group (labeled ${\em A}$) shows the state of the adsorbed
phase. The $Pt$ surface exhibits well defined propagating
fronts that correspond to conversion of the $Pt$ surface between the $hex$
and $1 \times 1$ phases. At the begining of the cycle shown (upper left) 
most of the surface is in the $hex$ phase (black) and there are islands of
the $1 \times 1$ phase (white). Since the $CO$ coverage is high the more
stable surface state is the $1 \times 1$ phase and the $1 \times 1$ 
islands quickly grow
and consume the $hex$ phase until the entire surface is $1 \times 1$ with
a few very small islands barely discernible in the top right panel of the
figure. With the surface in the $1 \times 1$ phase, $O$ adsorption is
high, $CO$ is consumed and its coverage falls. The $hex$ islands 
now grow and consume the $1 \times 1$ phase until the surface reverts 
to a nearly complete $hex$ structure with a few small $1
\times 1$ islands. If one observes the spatiotemporal dynamics over many
global oscillation periods one sees a constantly changing pattern of waves
of phase transformation sweeping over the surface but the dynamics is
sufficiently well synchronized that the cycling time between the ``$hex$-rich"
and ``$1 \times 1$-rich" phases is nearly constant. 

In the model the adsorbed $O$ atoms do not diffuse but $CO$ does. 
Apart from some small scale fluctuations the adsorbed species are
uniformly distributed and do not exhibit chemical waves (cf. ${\em A}$ 
panels in Fig.~\ref{spatd10r}). In this case
$D_{CO}$ is sufficiently large ($\delta=10$) that $CO$ diffusion is able
to maintain homogeneity over the finite simulation domain. A rough estimate
of the diffusion length is $\ell_D \sim (\delta \tau_{osc})^{1/2}$, where
$\tau_{osc}$ is the oscillation period. For the conditions of the
simulation described above we have $\tau_{osc}=640$,
$\ell_D \sim 80$ and $N=100$ so that the ratio of
the diffusion length to the system size is $\ell_D/L \sim 0.8$ and one
expects diffusion to homogenize the $CO$ concentration gradients, as is
indeed observed. The adsorbed species concentrations oscillate but there
is almost no pattern formation in the adsorbed layer.

\begin{figure}[htbp]
\begin{center}
\leavevmode
\epsfysize = 6in
\epsffile{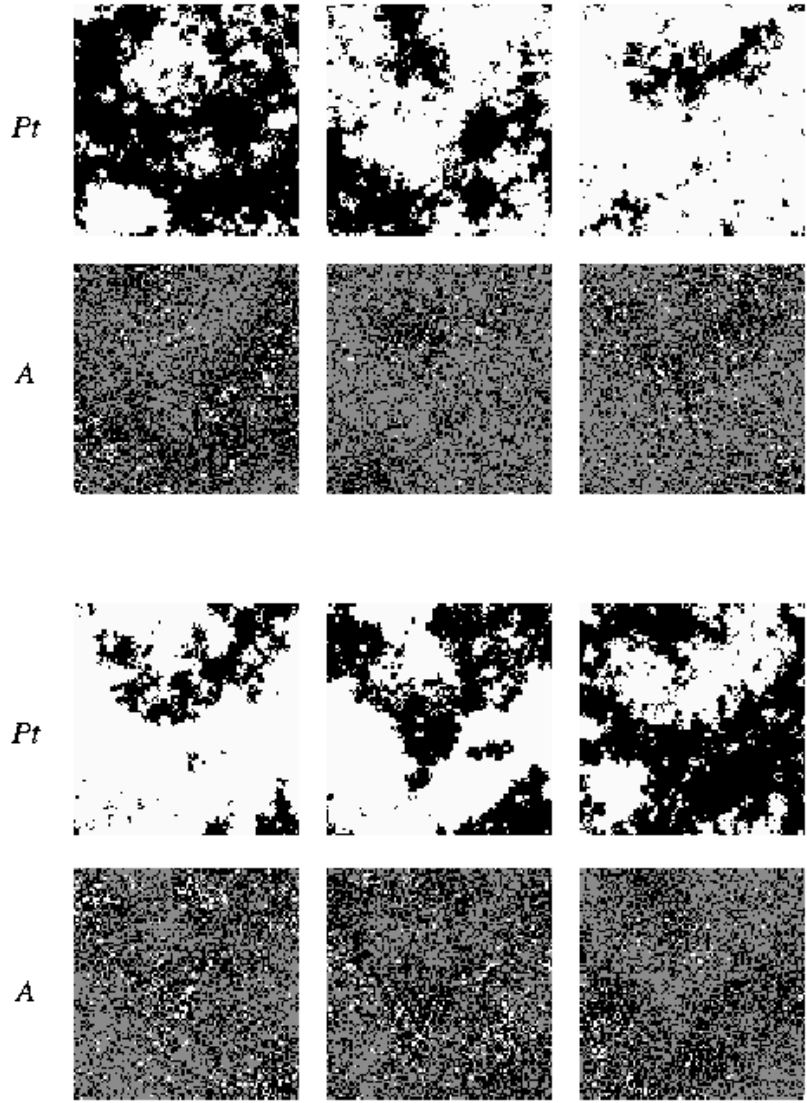}
\end{center}
\caption{Evolution of the $Pt(100)$ substrate and adsorbed phase in the 
regime of regular  relaxation oscillations for $k_2^{1 \times 1}=0.04$, 
$k_{1}^{1 \times 1}=0.017$ and $\delta=1/2$. Coding same as 
Fig.~\protect\ref{spatd10r}. Starting with time $t_1$ the
subsequent times are: $t_2=t_1+80$, $t_3=t_1+480$, $t_4=t_1+840$,
$t_5=t_1+960$ and $t_6=t_1+1110$.}
\label{spatd.5r}
\end{figure}

When $D_{CO}$ is smaller ($\delta=1/2$) there are time episodes where the 
waves of phase transformation that sweep the surface are no longer synchronized. 
This desynchronization is evident in the plots of $N_{hex}$ versus $t$ 
shown in Fig.~\ref{gosc.5} where one observes trajectory segments with 
large amplitude and others with small amplitude. In the time segments 
where the global concentrations have small amplitude 
the spatial structure is highly disorganized. 
Figure~\ref{spatd.5r} shows the spatial structure during such a 
small-amplitude episode where 
one sees that the irregular time variation of the global trajectory arises 
from the irregular surface structure: while waves of phase transformation still 
sweep across the $Pt$ surface, 
the nucleation and growth events have a highly random 
character. The surface does not revert to a nearly pure $1 \times 1$ or
$hex$ form since new nucleation and growth events occur before others are
completed. 
For the system parameters in this case we have $\tau_{osc} \sim 1100$ 
and $\ell_D \sim 24$ which is much smaller than the system length, 
$\ell_D/L \sim 0.24$. 
In this circumstance $CO$ diffusion will no longer be able to 
maintain spatial homogeneity in the concentration across the surface. This 
is evident in the lower panels (${\em A}$) in Fig.~\ref{spatd.5r} which show 
the local concentrations of the adsorbed species: there are now
significant spatial variations in the adsorbed species concentrations.  

Of course, the general structure of these results is to be 
anticipated on physical 
grounds. A fixed value of the diffusion coefficient and kinetic parameters implies a 
fixed diffusion length $\ell_D$. As the system size increases it will be 
impossible for diffusion to maintain a synchronized dynamics across the entire 
surface and one should observe phase turbulence. \cite{kuramoto} If the results 
for small diffusion are viewed on smaller length scales, small compared to 
the diffusion length, then synchronized dynamics will again be observed. The domain 
over which the global concentrations are computed in experiment is related to 
the area of the surface probed by the various measuring techniques. This point 
has been discussed in Refs.~\cite{moller1,moller2}.  Apart from these rather general 
observations it is interesting to recall that in the present model only $CO$ 
diffuses and it is the $Pt$ substrate state that exhibits the most well 
defined surface structure. This structure is best organized when there is 
no spatial pattern in the adsorbed phase. In this sense the standard arguments 
leading to phase turbulence take a somewhat different character that makes the 
results presented here less obvious.

\begin{figure}[htbp]
\begin{center}
\leavevmode
\epsfysize = 6in
\epsffile{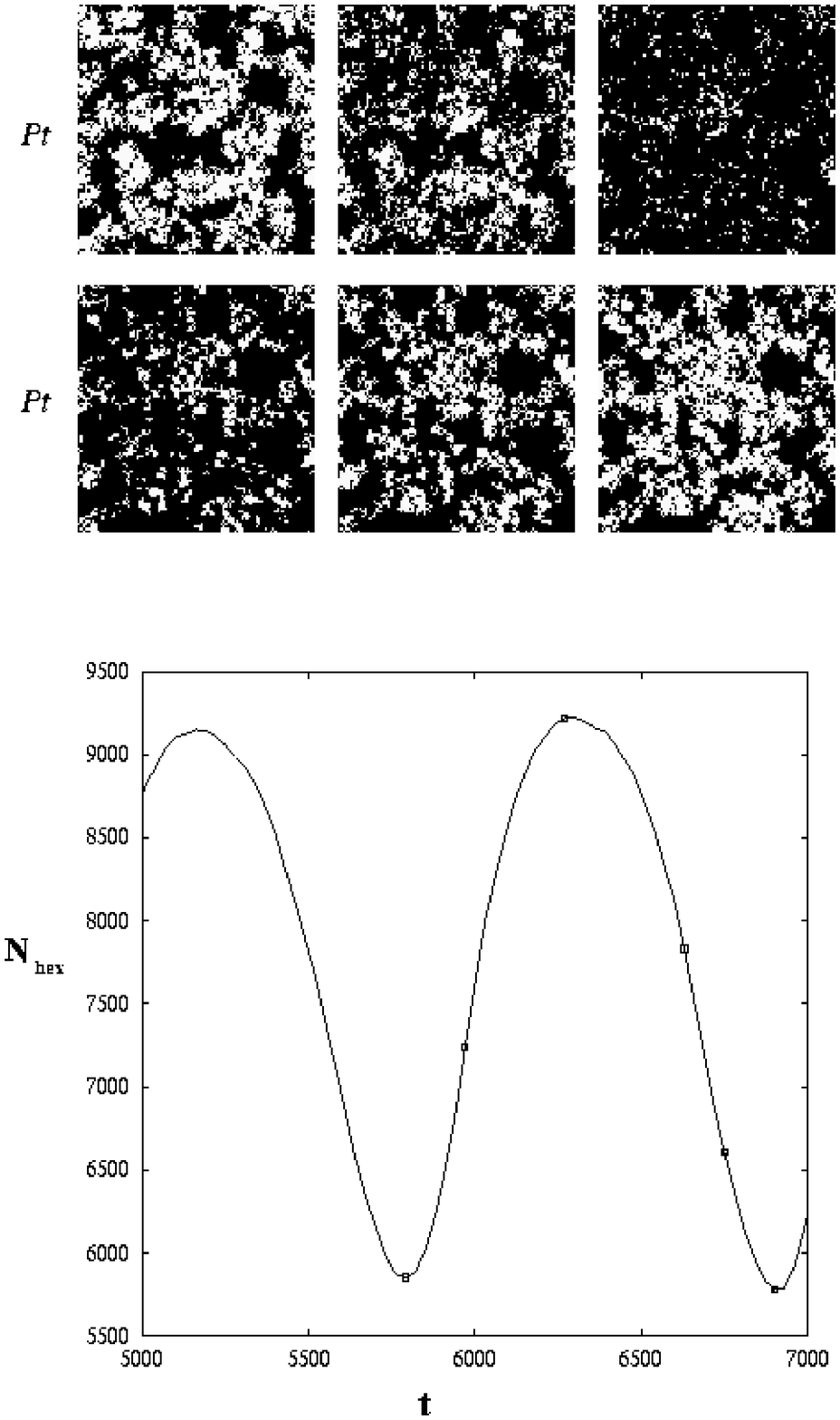}
\end{center}
\caption{Evolution of the $Pt(100)$ substrate in the regime of 
regular  harmonic oscillations for $k_2^{1 \times 1}=0.04$, 
$k_{1}^{1 \times 1}=0.005$ and $\delta=10$. Coding same as 
Fig.~\protect\ref{spatd10r}. The lower panel shows $N_{hex}$ versus 
$t$. The square points on the indicate the times at which the 
spatial structure of the $Pt$ surface is displayed in the upper 
panel.}
\label{shopf}
\end{figure}

Next, we consider the behavior near the bifurcation boundary where the oscillations 
are harmonic in character and appear to arise from a Hopf bifurcation. Here 
the situation is quite different, even for the large diffusion case where the 
global oscillations are well defined. In view of the fact that the global 
oscillations grow in amplitude and are small just within the bifurcation boundary, 
we need to examine how it is possible for such small-amplitude oscillations 
to arise in the spatially distributed medium. This can be understood from
a study of Fig.~\ref{shopf} 
which shows the state of the $Pt$ surface during one global oscillation 
period. The open squares on the curve showing $N_{hex}$ versus $t$ in 
lower panel of this figure 
indicate the times in the global oscillation cycle corresponding to the
$Pt$ surface structure in the upper panels. 
In contrast to the rapidly growing waves of phase transformation that sweep 
across the surface in the relaxation oscillation regime, here one observes that 
large portions of the $Pt$ surface remain in the $hex$ phase throughout the 
global oscillation cycle. The growth of $hex$ now largely occurs by repeated 
nucleation events which tend to ``fill in " the $1 \times 1$ seas. The $hex$ islands 
do, of course, grow and shrink to some extent but one does not observe the 
well-defined waves of phase transformation as in the global relaxation oscillation 
case. One sees that some $hex$ islands persist throughout the global
oscillation cycle and their shape changes slowly over several cycles. 
Consequently, a different spatial mechanism underlies these harmonic 
oscillations.

\section{Conclusion} \label{conc}
The automaton described in this paper incorporates the reactive 
processes in the adsorbed layer as well as phase transformations in the 
substrate. We have demonstrated that it is able to capture the 
essential of the features of catalytic $CO$ oxidation oscillations 
on $Pt(100)$. 
 
The main emphasis of this study was the exploration of the spatial
structure that underlies the global oscillatory dynamics. The 
nature of the global oscillations was found to depend 
sensitively on the coupling between the substrate phase transformations 
and the dynamics of the species in the adsorbed layer. This 
is evident from the variations in the oscillatory structure that 
occur when the diffusion coefficient of adsorbed $CO$ is varied. 
The oscillations are regular in character when $D_{CO}$ is 
large, or irregular, and eventually destroyed, when $D_{CO}$ is 
small. This behaviour is connected with the formation and 
synchronization 
of $hex$ and $1 \times 1$ islands on the substrate. In the case of large 
diffusion the $hex$ and $1 \times 1$ islands are large but the adsorbed
layer is disordered. When $CO$ diffusion is small
the islands are smaller in size but exist in the substrate and 
adsorbed layers. The dynamics in these two layers are strongly 
synchronized in this case. 

The form of the spatial structure in the $Pt$ surface also depends on
the harmonic or relaxational character of the global oscillations, which
can be tuned by varying $p_{CO}$, the partial pressure of $CO$. In the
relaxation oscillation regime well defined waves of phase transformation
sweep over the surface. One wave passes before new nucleation events begin
another cycle the phase transformaion process. Consequently one sees very
regular global oscillations. In the harmonic region, close to the 
bifurcation boundary, the global oscillations have small amplitude. As a
result not all the surface is converted to the $1 \times 1$ or $hex$
phases. There are no well defined waves of phase transformation, instead 
certain islands remain locked in one surface phase over several global 
cycles and nucleation events fill the ``seas" between the islands. A rich
surface structure underlies the global temporal evolution.

This automaton is a promising tool for study more complex surface 
catalytic oxidation reactions. Both the adsorbed layer kinetics 
and the nature of the surface reconstruction can be modified from that 
considered here. In particular, detailed studies of pattern formation 
at the mesoscopic level\cite{ertl4,ertl5} should be possible  
using such a model and the 
influence of fluctuations and the possible breakdown of deterministic 
descriptions can be considered.

\acknowledgments
The authors wish to thank professors G. Ertl and R. Imbihl
for fruitful discussions. The research of R.K. was supported in part by 
a grant from the Natural Sciences and Engineering Research Council of 
Canada.

\begin{thebibliography}{abracadabra}
\bibitem{pt110} The situation is different for $Pt(110)$ where global coupling 
through the gas phase is believed to play a significant role 
(cf. Ref.~\protect\cite{moller2}).
\bibitem{ptnote} Perhaps the simplest such model consists in assuming that 
phase transformation is proportional to the fractional occupancy of the 
neighborhood, in which case $F(n_{\alpha})=n_{\alpha}$,
but other forms cannot be excluded until all microscopic details of 
the phase transformation process are known. The form
$F(n_{\alpha})=n_{\alpha}^2$ used in this study gives a stronger weighting
to the local coverage and mimics some larger neighborhood rules. Both 
of these choices give well defined island structure and growth and 
qualitatively similar phase transformation dynamics. We have 
considered other forms as well, including $F(n_{\alpha})={\rm const}$ with 
restrictions on $k_{1H}$ and $k_{H1}$ which lead to less well defined 
surface structures.
\bibitem{binote} Such bifurcation diagrams do not have precisely defined
bifurcation boundaries in view of the probabilistic character of the 
dynamics. While individual realizations may not always conform to the
dynamics indicated by such a bifurcation diagram, in our case the
overwhelming majority of realizations do yield the behavior shown in the
figure.
\end{thebibliography}
\end{document}